\documentclass[twoside,final]{pro12}

\usepackage[utf8]{inputenc}
\usepackage{amsmath}
\usepackage{graphicx}
\usepackage{array}
\usepackage{SIunits}
\usepackage{lscape}
\usepackage{multirow}
\usepackage{float}
\usepackage{natbib}
\usepackage{hyperref}
\usepackage{textcmds}

\head{Normo et al.}{Fine structures in type II solar radio bursts }	

\begin{document}

\title{The location and propagation of fine structures in type II solar radio bursts}	
\author{S.~Normo\adress{\textsl Department of Physics and Astronomy, University of Turku, 20014, Turku, Finland}$\,\,
~^{\href{https://orcid.org/0009-0003-1737-4746} 
{\includegraphics[scale=0.005]{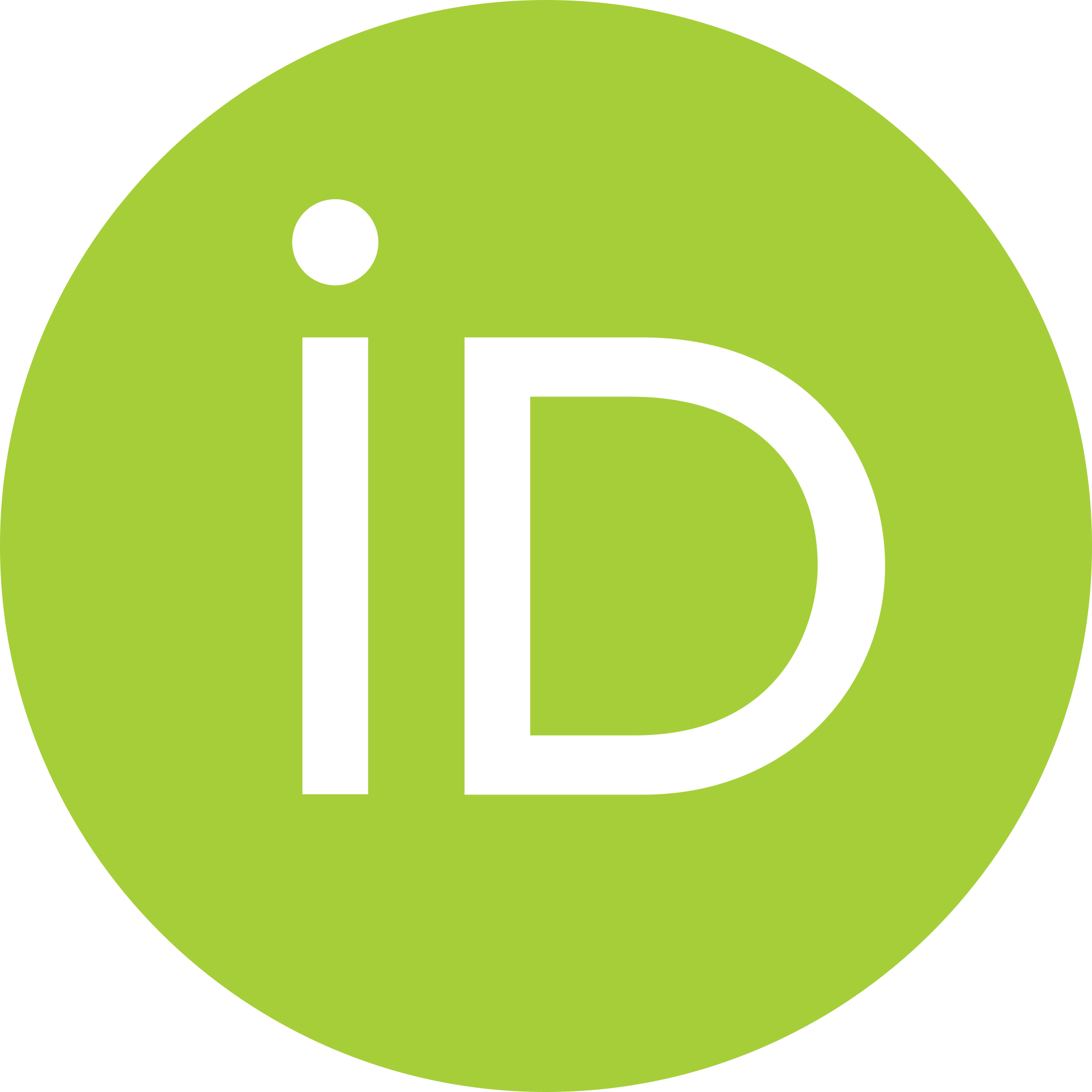}}}$,
K.~Bhandari$^1\,\,
~^{\href{https://orcid.org/0009-0002-0408-0354}
{\includegraphics[scale=0.005]{ORCID-icon.png}}}$,
A.~Nadiger$^1
~^{\href{https://orcid.org/0009-0009-0429-0010}
{\includegraphics[scale=0.005]{ORCID-icon.png}}}$,
and D.~E.~Morosan$^{1,}$\adress{\textsl Turku Collegium for Science, Medicine and Technology, University of Turku, 20014, Turku, Finland}$\,\,
^*
~^{\href{https://orcid.org/0000-0002-8416-1375}
{\includegraphics[scale=0.005]{ORCID-icon.png}}}$\\
\footnotesize $^*$Corresponding author: \href{mailto:diana.morosan@utu.fi}{diana.morosan@utu.fi}}\normalsize

\maketitle

\footnotesize \textit{Citation:}\\
Normo et al. (2026). The location and propagation of fine structures in type II solar radio bursts, in \textit{Planetary, Solar and Heliospheric Radio Emissions X. L. Lamy, C. K. Louis, G. Fischer, D. Morosan, P. Zarka eds. OSU Pyth\'{e}as/AMU, Observatoire de Paris}. Preprint. \href{https://doi.org/10.25935/prex-tadg}{doi:10.25935/prex-tadg}

\begin{abstract}
Solar eruptions such as coronal mass ejections can drive collisionless shocks that are good particle accelerators. Electrons accelerated by these shocks can be observed remotely via the electromagnetic emission they generate at low radio frequencies. The radio signatures of shock-accelerated electrons at the Sun are type II radio bursts that can be used to track the propagation of the shock wave in the solar corona and beyond. However, type II radio bursts can have complex morphologies in dynamic spectra, being composed of numerous fine time and frequency structures. Here, we aim to determine the location and propagation of the fine structures composing type II bursts using radio imaging from the Nan\c cay Radioheliograph. We investigate the origin of a type II radio burst that was only co-temporal with a flare and a coronal wave, and it was not associated with a CME eruption. The type II burst still showed complex morphology. We find that emission lanes and fine structures composing the type II burst originate from multiple locations around the flare site. The source regions also move in peculiar non-uniform propagation directions following the shock expansion. Our findings are consistent with the idea that multiple radio emission source regions form as a shock propagates through the solar corona.
\end{abstract}

\section{Introduction}

Various processes in the solar corona accelerate electrons that emit across the electromagnetic spectrum. In the radio domain, solar activity can be associated with solar radio bursts of types I--V. These bursts are identified in dynamic spectra where they can exhibit a variety of fine structures in addition to their general spectral appearance. Radio signatures of electrons accelerated at coronal shock waves at the Sun are type II radio bursts observed in the metric and decametric wavelengths \citep[][]{ne85,mann1995,ma96}. In a dynamic spectrum, type II bursts are identified as emission lanes drifting slowly from higher to lower frequencies at the fundamental and/or harmonic of the plasma frequency. Type II bursts are produced via the plasma emission mechanism \citep{Melrose1985} where accelerated electron beams excite Langmuir waves and eventually some of the energy of the Langmuir waves is converted into radio emission. Type II bursts are usually associated with expanding shocks in the corona as the radio sources are co-spatial with the passage of associated CMEs and/or associated shock waves \citep[e.g.][]{zu18, mancuso19, Morosan2020a}. Recent imaging observations of type II bursts with complex morphologies that deviate from a two-lane fundamental-harmonic structure revealed multiple radio sources at spatially separated regions \citep[e.g.][]{mo19a,Zhang2024a}. This is attributed to multiple hot spots for electron acceleration at different regions of the shock \citep[e.g.][]{mo19a, morosan2024}.

The fine structures in type II bursts can be split into two categories: the presence of multiple emission lanes composing a type II burst in dynamic spectra and the fine-structured bursts composing these type II emission lanes. Type II multi-lanes are a phenomenon where a type II burst consists of two or more lanes, that may have similar drift rates but not always the same spectral morphology \citep[][]{zimovets2015, alissandrakis2021}. When multi-lanes have a similar spectral morphology, they are known as split-bands \citep[e.g.][]{vr01}. Given the somewhat similar characteristics of these lanes in dynamic spectra, they are likely generated in regions of the shock that are in close proximity to each other. Recent radio imaging studies have indeed demonstrated that multi-lanes in type II bursts, including split bands, originate from different but neighbouring regions at the shock \citep[e.g.][]{bhunia2023, Morosan2023, normo2025}. In addition, each of these lanes are composed of numerous fine structures, many of them documented in \citet[][]{magdalenic20}. The most well-known fine-structured bursts are called `herringbones' and are characterised by short-duration drifting bursts towards either low or high frequencies stemming from a type II `backbone' \citep{ho83,ca87}, but they can also occur on their own \citep[][]{ho83,mo19a}. Herringbone bursts represent signatures of individual electron beams accelerated by a shock predominantly at the flanks of a CME and which expands through the corona in a quasi-perpendicular direction to the surrounding magnetic field \citep{zl93, ca13, mann2018, mo19a}. 

In this paper, we investigate the location and evolution of different spectral features composing a complex type II radio burst, which occurred in the absence of a CME.

\section{Observations and data analysis}

\subsection{Radio observations}

\begin{figure}[t]
\centering
\includegraphics[width=\linewidth]{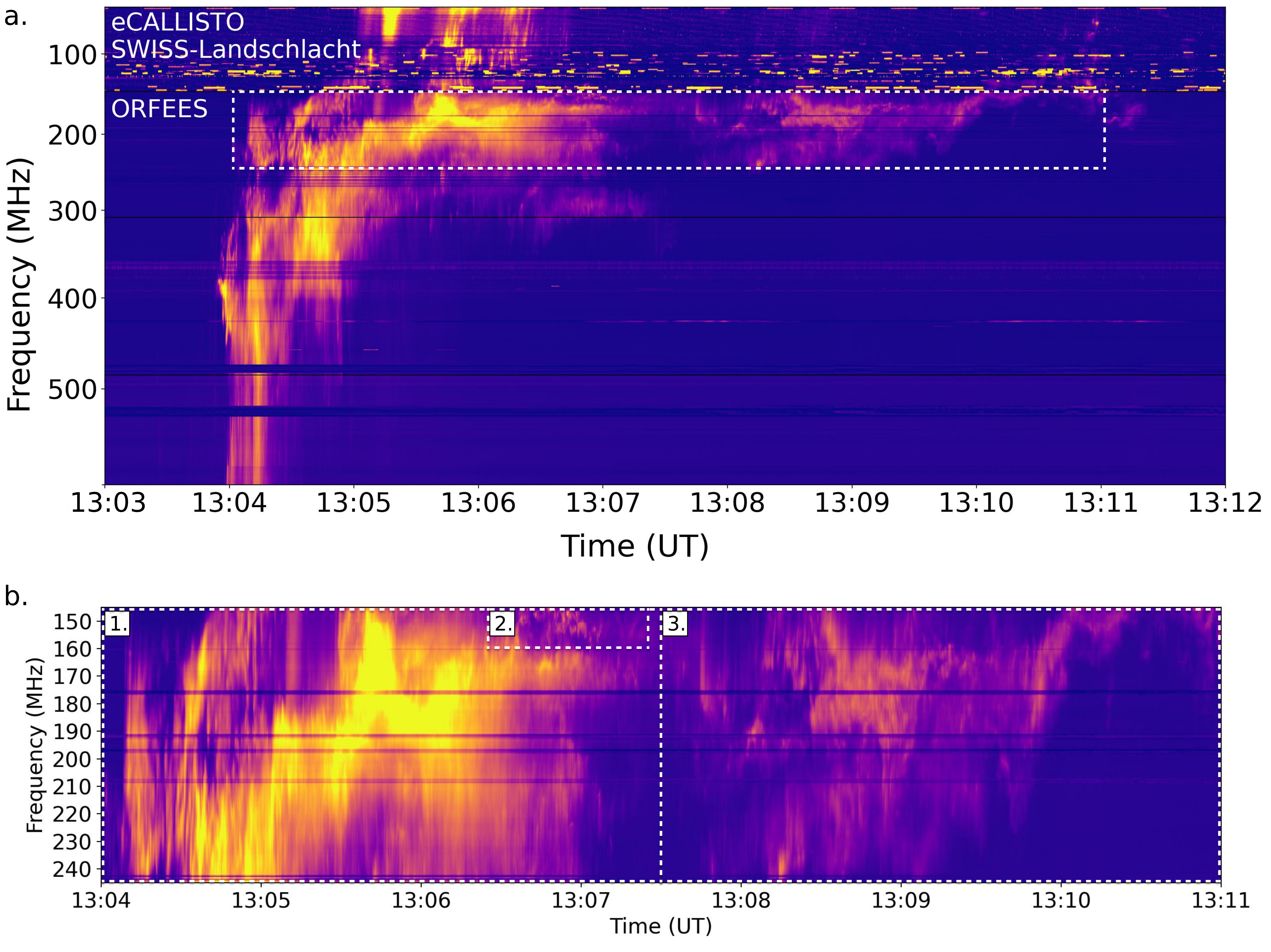}
\caption{Dynamic spectra of the type II burst. (a) Combined dynamic spectrum at $50-600 \, \mathrm{MHz}$ from 13:03 to 13:12 UT observed by e-CALLISTO SWISS-Landschlacht and ORFEES. The white dashed rectangle outlines the part of the spectrum shown in panel (b). (b) Zoomed-in dynamic spectrum at $145-245 \, \mathrm{MHz}$ from 13:04 to 13:11 UT. The white dashed rectangles outline the three parts of the spectrum studied in more detail.}
\label{fig:1}
\end{figure}

A type II solar radio burst was observed by multiple radio instruments on 11 March 2025 between approximately 13:04 and 13:11 UT. Figure \ref{fig:1} shows the dynamic spectra of the type II as observed by the e-CALLISTO \citep{Benz2005,Benz2009} SWISS-Landschlacht spectrometer and by the ORFEES \citep[Observation Radio pour FEDOME et l'\'Etude des \'Eruptions Solaires;][]{Hamini2021} radio spectrograph. The spectrum in Figure \ref{fig:1}a covers a frequency range of $50-600 \, \mathrm{MHz}$ between 13:03 and 13:12 UT. Figure \ref{fig:1}b shows a zoomed-in part of the dynamic spectrum observed by ORFEES at $145-245 \, \mathrm{MHz}$ outlined with a white dashed rectangle in Figure \ref{fig:1}a. In this study, we are focusing on three different parts of the spectrum, boxed and labelled in Figure \ref{fig:1}b, where different fine structures and emission lanes are observed. The first part (Box 1) shows a multitude of herringbones composing multiple lanes of the type II burst, some of which show a bright wavy backbone. The second part (Box 2) shows a short-duration reverse-drifting structure around $150 \, \mathrm{MHz}$. The third part (Box 3) shows a non-drifting fainter type II lane composed of herringbones. Interferometric imaging of the type II burst is available from the Nan\c cay Radioheliograph \citep[NRH;][]{Kerdraon1997} at several frequencies between $150$ and $450 \, \mathrm{MHz}$. Here we use NRH imaging at $150.9 \, \mathrm{MHz}$, $173.2 \, \mathrm{MHz}$ and $228.0 \, \mathrm{MHz}$. The temporal resolution of the imaging is $0.25 \, \mathrm{s}$. From the radio imaging observations, we extract the centroids of the radio sources by fitting 2D elliptical Gaussian functions to the radio images \citep[see for example the methods of][]{mo19a}. We do this fitting at each imaging time step within the time periods of interest to study how the radio sources evolve in space and in time at the selected frequencies.

\subsection{EUV observations}

 To investigate the solar activity accompanying the type II burst, we use extreme ultraviolet (EUV) observations of the Sun from the Atmospheric Imaging Assembly \citep[AIA;][]{le12} on board the Solar Dynamics Observatory \citep[SDO;][]{pe12}. The type II is associated with an M1.1-class flare originating from the NOAA AR 14024 situated at that moment close to the west solar limb. No evidence of a CME eruption was found in EUV or white-light images. White-light images captured by the Large Angle Spectroscopic Coronagraph \citep[LASCO;][]{br95} on board the Solar and Heliospheric Observatory \citep[SOHO;][]{do95} show a jet-like structure propagating away from the Sun. However, this jet originates from a different active region than the flare and is launched after the end of the type II event. Thus, it is not associated with the type II burst. Only a prominent EUV wave can be observed in EUV running difference images propagating away from the flaring region. This EUV wave is probably a fast-mode wave that propagates in the low solar corona \citep{Uchida1968,Mann1999,long2008,Warmuth2015,Vrsnak2016} and given the absence of a CME, it may be a freely propagating wave associated with the flare. Figures \ref{fig:2}b, \ref{fig:3}a and \ref{fig:4}b show the evolution of the flare and EUV wave in running difference images from the AIA $211 \, \mathrm{\AA}$ passband. The radio centroids obtained from the NRH imaging are also included in these figures. The centroids are colour-coded through time corresponding to the same colouring of the time steps shown in Figures \ref{fig:2}a, \ref{fig:3}b and \ref{fig:4}a. The type II burst is already present during the onset of the EUV wave propagation (Figure~\ref{fig:2}). 

\section{Results}

\begin{figure}[t]
\centering
\includegraphics[width=0.9\linewidth]{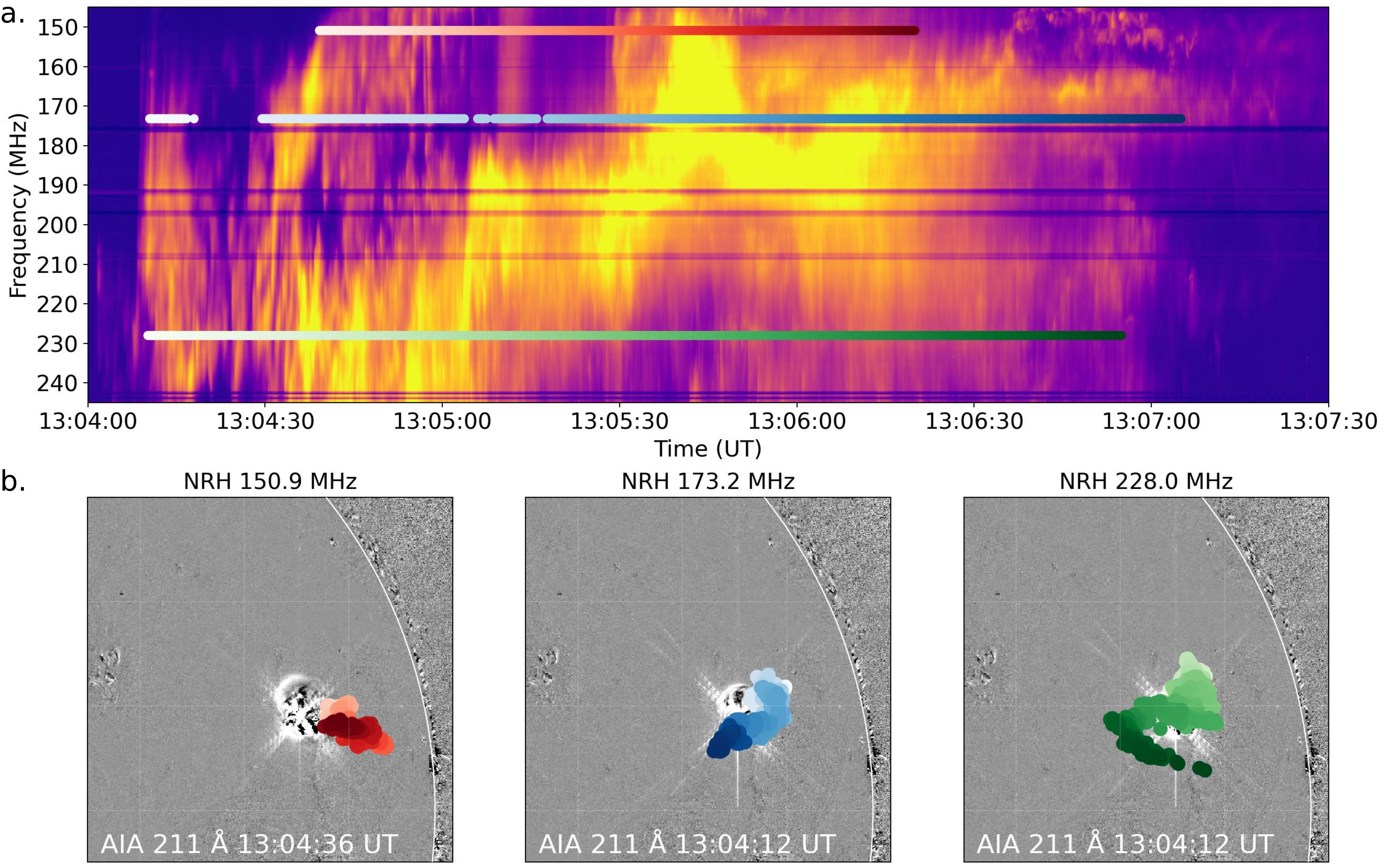}
\caption{Dynamic spectrum and radio centroids of Box 1 outlined in Figure \ref{fig:1}b. (a) Zoomed-in dynamic spectrum between $145-245 \, \mathrm{MHz}$ at 13:04:00--13:07:30 UT. (b) NRH centroids at $150.9 \, \mathrm{MHz}$ (reds), $173.2 \, \mathrm{MHz}$ (blues) and $228.0 \, \mathrm{MHz}$ (greens) overlaid on SDO/AIA running difference images at $211 \, \mathrm{\AA}$. The time of the AIA images are chosen to be close to the starting time of the centroids at each frequency.}
\label{fig:2}
\end{figure}

The radio images show that during the three parts of the spectrum that were identified in the previous section, the fine-structures composing the type II emission originate from different locations (see the centroids in Figures \ref{fig:2}--\ref{fig:4}). There are three possible regions where the shock accelerates electrons to produce these fine structures, as outlined below.

\subsection{Box 1}

The part of the spectrum outlined in Box 1, spanning from 13:04:00 to 13:07:30 UT at $145-245 \, \mathrm{MHz}$, is characterised by multiple type II lanes composed of herringbones, one of these exhibiting a bright wavy backbone (Figure \ref{fig:2}a). Figure \ref{fig:2}b shows the radio centroids at $150.9 \, \mathrm{MHz}$ (reds), $173.2 \, \mathrm{MHz}$ (blues) and $228.0 \, \mathrm{MHz}$ (greens). The plane-of-sky source locations of the centroids are in regions west and north-west of the flare site, along the observed EUV wave. The centroids at $150.9 \, \mathrm{MHz}$ are tracked for $1 \, \mathrm{min} \ 41 \, \mathrm{s}$ from 13:04:39 UT onward. Within this time period, the centroids show very slight movement to the west and back. The centroids at $173.2 \, \mathrm{MHz}$ and $228.0 \, \mathrm{MHz}$ are tracked for a longer time period starting already at 13:04:10 UT for $2 \, \mathrm{min} \ 55 \, \mathrm{s}$ and $2 \, \mathrm{min} \ 45 \, \mathrm{s}$, respectively. The centroids at both frequencies originate from a similar location north-west of the flare. The radio sources then cross the flare site and propagate toward regions east of the flare. The $173.2 \, \mathrm{MHz}$ and $228.0 \, \mathrm{MHz}$ centroids have a similar curved path around the flaring region, but the $228.0 \, \mathrm{MHz}$ centroids propagate even farther towards the south. This shows that already during the onset of the EUV wave, the shock was able to accelerate electrons close to the flare location. 

\subsection{Box 2}

\begin{figure}[t]
    \centering
    \includegraphics[width=0.8\linewidth]{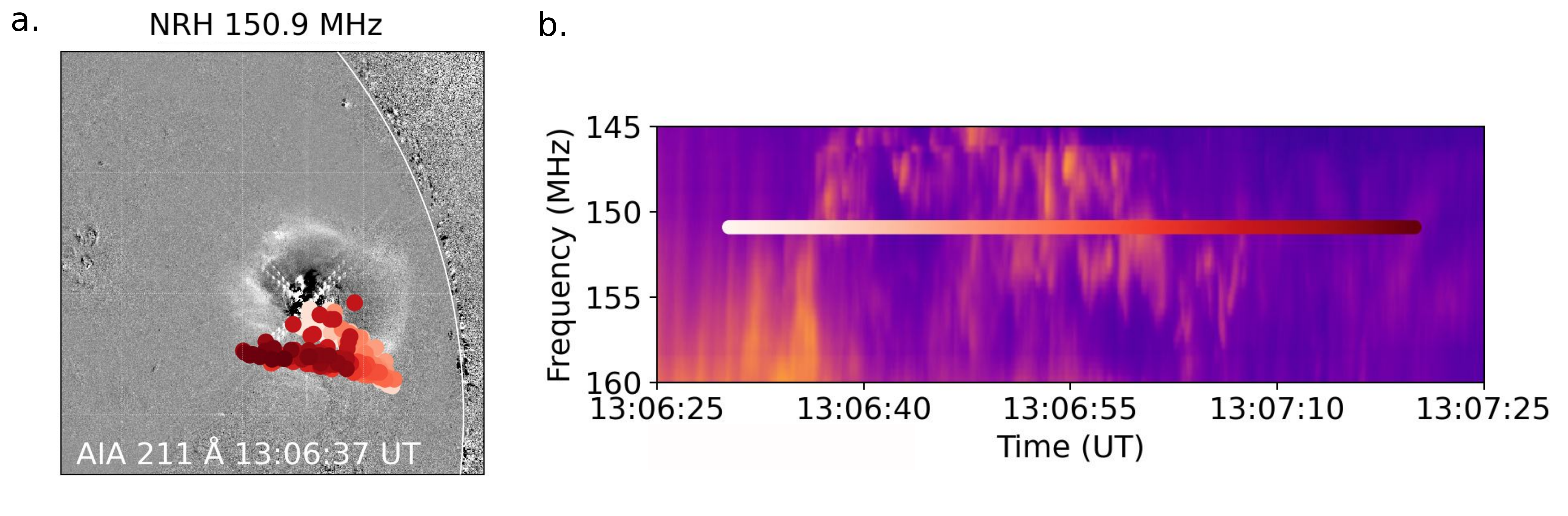}
    \caption{Dynamic spectrum and radio centroids of Box 2 outlined in Figure \ref{fig:1}b. (a) NRH centroids at $150.9 \, \mathrm{MHz}$ (reds) overlaid on SDO/AIA running difference images at $211 \, \mathrm{\AA}$ at 13:06:37 UT. (b) Zoomed-in dynamic spectrum between $145-160 \, \mathrm{MHz}$ at 13:06:25--13:07:25 UT.}
    \label{fig:3}
\end{figure}

The part of the spectrum outlined in Box 2, at 13:06:25--13:07:25 UT between $145 \, \mathrm{MHz}$ and $160 \, \mathrm{MHz}$, was selected due to the presence of a lane that differs in morphology from the lanes in Box 1. The bursts composing this lane have a smaller bandwidth than the herringbones in Box 1 (Figure \ref{fig:3}b). The centroids (reds) in Figure~\ref{fig:3}a correspond to a frequency of $150.9 \, \mathrm{MHz}$ and span over an interval of $50 \, \mathrm{s}$ starting at 13:06:30~UT to cover the full extent of the type II lane composed of smaller bandwidth bursts. The starting location of these centroids is close to that of the herringbones in Box 1 and overlaps the centre of the eruption in the plane-of-sky images. The centroids have an overall \q{zig-zag} movement within the tracked time period, but there may also be some overlap with the fainter sources of herringbones in Box 1. The centroids first move farther away from the eruption toward south-west as the type II lane crosses $150.9 \, \mathrm{MHz}$ in the spectrum. As this lane ends, the centroids move back to a location east of the flare site. The location is similar to where the centroids at $173.2 \, \mathrm{MHz}$ and $228.0 \, \mathrm{MHz}$ stopped in Box 1. These positions suggest that the radio sources composing this lane are in a different location than the earlier radio sources.

\subsection{Box 3}

\begin{figure}[t]
    \centering
    \includegraphics[width=0.9\linewidth]{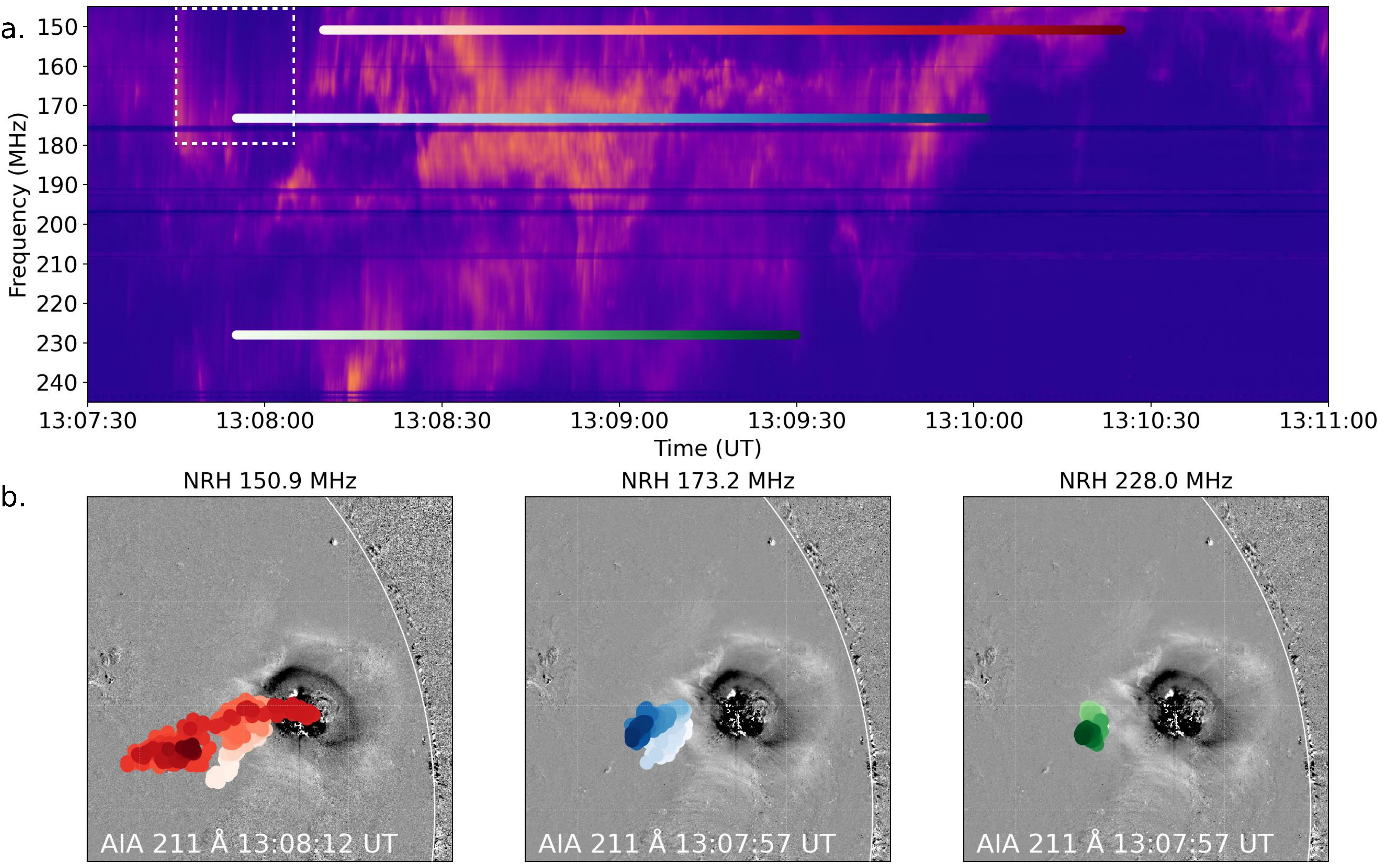}
    \caption{Dynamic spectrum and radio centroids of Box 3 outlined in Figure \ref{fig:1}b. (a) Zoomed-in dynamic spectrum between $145-245 \, \mathrm{MHz}$ at 13:07:30--13:11:00 UT. The white dashed rectangle outlines the part of the spectrum shown in Figure \ref{fig:5}b. (b) NRH centroids at $150.9 \, \mathrm{MHz}$ (reds), $173.2 \, \mathrm{MHz}$ (blues) and $228.0 \, \mathrm{MHz}$ (greens) overlaid on SDO/AIA running difference images at $211 \, \mathrm{\AA}$. The time of the AIA images is chosen to be close to the starting time of the centroids at each frequency.}
    \label{fig:4}
\end{figure}

\begin{figure}
    \centering
    \includegraphics[width=0.9\linewidth]{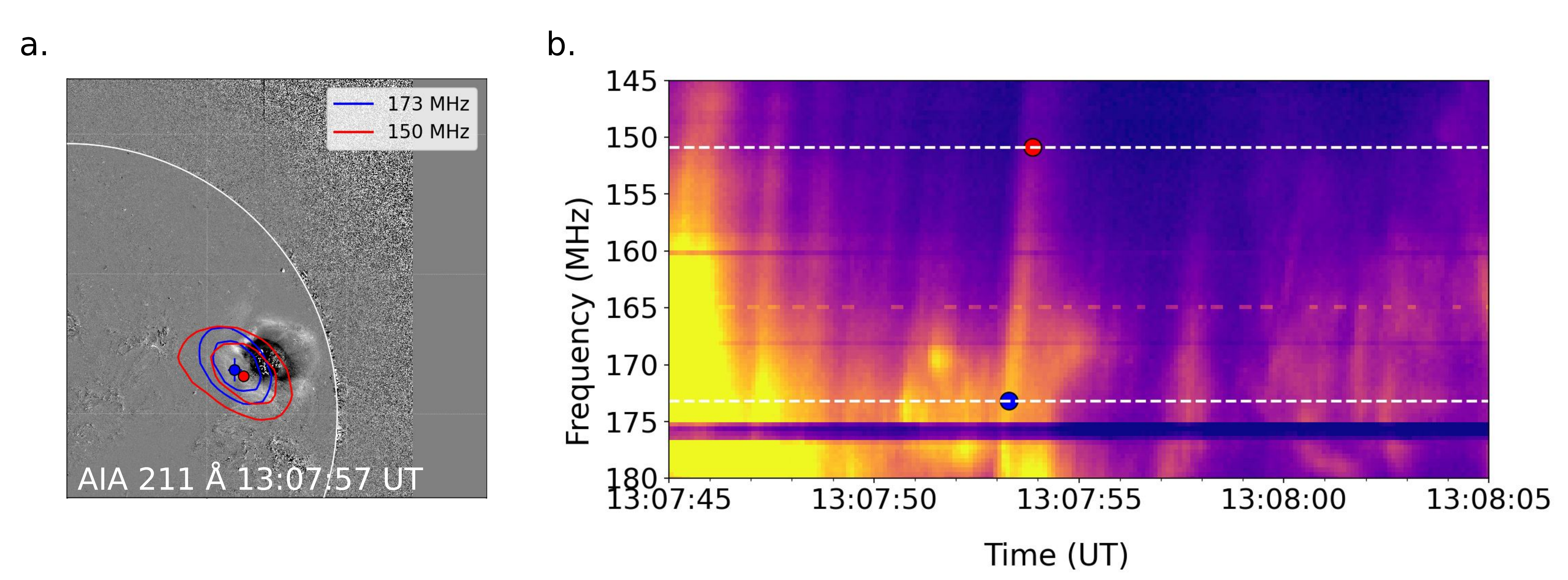}
    \caption{Dynamic spectrum and radio centroids of a single herringbone burst at two frequencies. (a) NRH centroids at $150.9 \, \mathrm{MHz}$ (red) and $173.2 \, \mathrm{MHz}$ (blue) overlaid on SDO/AIA running difference images at $211 \, \mathrm{\AA}$. (b) Zoomed-in dynamic spectrum outlined in Figure \ref{fig:4}a between $145-180 \, \mathrm{MHz}$ at 13:07:45--13:08:05 UT. }
    \label{fig:5}
\end{figure}

The part of the spectrum outlined in Box 3, at 13:07:00 to 13:11:00 UT between $145 \, \mathrm{MHz}$ and $245 \, \mathrm{MHz}$, shows faint emission lanes composed of herringbones. These emission lanes do not in general show a frequency drift. Figure \ref{fig:4}a shows a dynamic spectrum of the lanes during 13:07:30--13:11:00 UT. The centroids at $150.9 \, \mathrm{MHz}$ (reds) were tracked for $2 \, \mathrm{min} \ 15 \, \mathrm{s}$ starting from 13:08:10 UT. The centroids at $173.2 \, \mathrm{MHz}$ (blues) and $228.0 \, \mathrm{MHz}$ (greens) were tracked for $2 \, \mathrm{min} \ 7 \, \mathrm{s}$ and $1 \, \mathrm{min} \ 35 \, \mathrm{s}$, respectively, from 13:07:55 UT onward. The starting locations of the centroids at each frequency are in eastern regions compared to the flare location, as can be seen in Figure \ref{fig:4}b. The centroids at $150.9 \, \mathrm{MHz}$ and at $173.2 \, \mathrm{MHz}$ propagate along an arch-like path first toward the eruption and then back away from it. The centroids at $228.0 \, \mathrm{MHz}$ show very little movement around the region where they originate. This time period shows that the electron acceleration regions are farther away from the flare site compared to earlier times, likely due to the expansion of the shock in the corona.

The three time and frequency periods from the dynamic spectrum, corresponding to the three boxes defined above, show an overall movement of the source regions of accelerated electrons as the shock expands in the corona. The single-frequency centroids show the locations of the many fine structures composing the different lanes of the type II burst. However, tracking just one individual fine structure across multiple frequencies is challenging due to NRH's sparse frequency coverage. An example of an individual fine structure that can be tracked across frequency is in Figure \ref{fig:5}a. Figure \ref{fig:5}a shows the radio contours and centroids at $173.2 \, \mathrm{MHz}$ (blue) and at $150.9 \, \mathrm{MHz}$ (red) corresponding to the frequency drift of a single herringbone. The herringbone is situated in the eastern regions of the flare site, propagating away from this region. The overall location is in the same region as all centroids in Box 3. 

\section{Discussion and conclusions}

In this study, we used NRH imaging to investigate the location and propagation of fine structures composing a complex type II solar radio burst which occurred in the absence of a CME. The type II was associated with a strong flare and a prominent coronal wave. A CME is not always needed to drive a coronal shock since type IIs have been previously observed in their absence \citep[e.g.][]{Magdalenic2012,Su2015,Kumar2016,Morosan2023}. Suggestions for alternative mechanisms for type II burst generation, other than a CME-driven shock, include a blast wave ignited by the pressure pulse of a flare \citep[e.g.][]{Magdalenic2010,Magdalenic2012,Kumar2016} or the bulk plasma motion acting as a temporary driver launching a freely propagating wave that steepens into a shock \citep[e.g.][]{Su2015,Morosan2023}. For example, the generation of a type II without an associated CME was attributed to the steepening of a coronal wave into a shock \citep{Veronig2010,Warmuth2011,Mann2022,Mann2023} in low Alfv\'en speed regions, where the coronal wave was launched by a failed eruption \citep{Morosan2023}. In the present study, we observe a complex type II burst that is composed of emission originating from multiple sources close to the EUV wave. Since we also observe that some of the radio centroids that are projected on the two-dimensional EUV image reside over the flaring region (see e.g. Figures \ref{fig:2}, \ref{fig:3} and \ref{fig:4}), it is possible that they are in the three dimensional space situated on a shock wave dome. However, due to the lack of multi-viewpoint observations in EUV, it is difficult to determine the geometry of the shock. Images in the AIA $304\ \mathrm{\AA}$ passband show plasma movement after the flare onset, possibly suggesting a failed eruption over the active region after the launch of the wave. Although during the EUV wave onset, it is difficult to observe any bulk plasma movement, which would indicate a failed eruption, due to the high brightness of the flare. The acceleration mechanism of the type II generating electrons is likely similar to the scenario proposed in \citet{Morosan2023}. In the present study, the conditions of the ambient solar corona are such that the coronal wave can steepen into a shock at more than one location since we observe the emission to originate from multiple spatially separated locations. The positions of the type II burst sources are then a good indication of where this steepening occurs relative to the eruption region. Previous studies have found that favourable conditions for type II generation are, for example, the coronal wave or the CME interacting with a coronal streamer \cite[e.g.][]{Cho2008,Kong2015,kouloumvakos21,Morosan2023}. In these regions of closed magnetic field structures and high density, low Alfvén speed and quasi-perpendicular shock geometry can be achieved. The multiple source locations are also consistent with previous studies, which found that type II producing electrons can be accelerated at various locations on the surface of a shock \citep[e.g.][]{mo19a,Zucca2025}. In addition, the observed type II is composed of numerous herringbone structures, which \cite{Mann2022} found to be efficiently produced by shocks with a nearly perpendicular geometry and an Alfvén Mach number of approximately two.

The complex morphology of the type II burst and the multiple source locations suggest that the observed emission maps to different parts of the shock surface as well as slightly different plasma conditions. The bright central part of the type II burst, which is sometimes called a `backbone', also has a non-uniform drift. This non-uniform drift may be related to the non-radial movement that we see in the type II centroids. However, this cannot be fully understood using single-frequency imaging, and it requires imaging capabilities over a large frequency range to track the position of the non-uniform drift. A study investigating a spectral bump in an otherwise uniformly drifting band-split type II burst found that the separation of sources corresponding to each split band increased during this bump and then decreased again after the frequency bump \citep{Zhang2024b}. The non-uniform drifts and spectral bumps present in type II bursts are indicative of density or magnetic field inhomogeneities in the ambient plasma. Future studies of complex type II bursts that combine radio imaging over a large frequency range with magnetohydrodynamic modelling of the solar corona can provide new insights into the plasma properties of the source regions of the observed type II emission lanes and fine structures.

\section*{Acknowledgements}

 All authors acknowledge the Research Council of Finland project `SolShocks' (grant number 354409). S.N. acknowledges the Vilho, Yrj\"o and Kalle V\"ais\"al\"a Foundation of the Finnish Academy of Science and Letters. This study has received funding from the European Union's Horizon Europe research and innovation programme under grant agreement No.\ 101134999 (SOLER). We thank the Radio Solar Database service at LESIA \& USN (Observatoire de Paris) for making the NRH/ORFEES data available. We also thank the eCALLISTO network for the availability of radio spectra.

 \section*{Data availability statement}

 The NRH dataset is freely available from the Solar Radio Database @ Nan{\c c}ay website: \url{https://rsdb.obs-nancay.fr}. The AIA images are available from the Joint Science Operations Center (JSOC): \url{http://jsoc.stanford.edu}.

\newcommand{\newblock}{}
\bibliographystyle{mnras}
\bibliography{bibliography}

\end{document}